\documentclass{emulateapj}

\newcommand{\lta}{{\>\rlap{\raise2pt\hbox{$<$}}\lower3pt\hbox{$\sim$}\>}}
\newcommand{\gta}{{\>\rlap{\raise2pt\hbox{$>$}}\lower3pt\hbox{$\sim$}\>}}

\shorttitle{Clustering around SDSS J1030+0524}
\shortauthors{Stiavelli et al.}
\slugcomment{Accepted by ApJ Letters}

\begin{document}

\title{Evidence for Primordial Clustering \\
       Around the QSO SDSS J1030+0524 at \lowercase{z}=6.28\footnote{Based, in part, on data obtained at the W.M.\ Keck Observatory, which is operated as a scientific partnership between the California Institute of Technology, the University of California, and NASA, and was made possible by the generous financial support of the W.M.\ Keck Foundation.}}

\author{
M.~Stiavelli\altaffilmark{1}, 
S.~G.~Djorgovski\altaffilmark{2},
C.~Pavlovsky\altaffilmark{1},
C.~Scarlata\altaffilmark{1,3},
D.~Stern\altaffilmark{4},
A.~Mahabal\altaffilmark{2},
D.~Thompson\altaffilmark{2},
M.~Dickinson\altaffilmark{5},
N.~Panagia\altaffilmark{6},
G.~Meylan\altaffilmark{7}
}

\email{mstiavel@stsci.edu, george@astro.caltech.edu, cherylp@stsci.edu,
       scarlata@phys.ethz.ch, stern@zwolfkinder.jpl. nasa.gov, aam@astro.caltech.edu,
       djt@irastro.caltech.edu, med@ noao.edu, panagia@stsci.edu,
       georges.meylan@epfl.ch}

\altaffiltext{1}{Space Telescope Science Institute,
                 3700 San Martin Drive, Baltimore, MD 21218}
                 
\altaffiltext{2}{California Institute of Technology,
                 MS 105-24, Pasadena, California 91125}
                 
\altaffiltext{3}{Present address:  Dept.\ of Physics,
                 Swiss Federal Institute of Technology (ETH-Zurich),
                 ETH H{\"o}nggerberg, CH-8093, Zurich, Switzerland}
                 
\altaffiltext{4}{Jet Propulsion Laboratory,
                 Mail Stop 169-506, Pasadena, CA 91109}
                 
\altaffiltext{5}{National Optical Astronomical Observatories,
                 P.O.\ Box 26732, Tucson, AZ 85726}

\altaffiltext{6}{ESA Space Telescope Division,
                 3700 San Martin Drive, Baltimore, MD 21218}
                                  
\altaffiltext{7}{Laboratoire d'astrophysique,
                 Ecole Polytechnique F\'ed\'erale de Lausanne, Observatoire,
                 CH-1290 Chavannes-des-Bois, Switzerland}

\begin{abstract}
We present tentative evidence for primordial clustering, manifested as an
excess of color-selected objects in the field of the QSO SDSS J1030+0524 at
redshift $z = 6.28$. 
We have selected objects red in $i_{775}-z_{850}$ on the basis of Hubble Space
Telescope Advanced Camera for Surveys imaging of a field centered on
the QSO.  Compared to data at
comparable depth obtained by the GOODS survey, we find an excess of
objects with $(i_{775}-z_{850}) \geq 1.5$ in the QSO field.
The significance of the detection is
estimated to be $\sim$97\% on the basis of the counts alone and
increases to 99.4\% if one takes into account the color distribution.
If confirmed this would represent the highest redshift example
of galaxy clustering and would have implications on models for the growth
of structure. Bias-driven clustering of first luminous objects forming
in the highest peaks of the primordial density field is expected in most
models of early structure formation. The redshift of one of the candidates 
has been found to be $z=5.970$ by our spectroscopy with Keck I/LRIS, confirming
the validity of our color selection.
\end{abstract}


\section{Introduction}

Exploration of the early structure formation and the reionization era
is now one of the most active areas of cosmology; see, e.g., 
Madau (2000), Loeb \& Barkana (2001), Barkana \& Loeb (2001),
Stiavelli (2002), or Djorgovski (2004) for recent reviews and references.
The first hints of an approach to reionization, signaled by the dramatic
increase in the intergalactic hydrogen absorption at $z \simeq 6$, was
detected in the spectra of high-$z$ QSOs (Becker et~al.\ 2001, 
Djorgovski et~al.\ 2001, White et~al.\ 2003, etc.).
In particular, there was the possible detection of an extended, optically
thick absorption due to neutral hydrogen at $\lambda_{rest} < 1216$~\AA\ 
(Gunn \& Peterson 1965) in the spectrum of QSO SDSS J1030+0524 at
$z = 6.28$ (Fan et~al.\ 2001).  These results strongly suggest
the end of the cosmic reionization at $z \approx 6$.  However, WMAP's
tentative detection of a large amplitude signal in the temperature-polarization 
maps of the cosmic microwave background (Kogut et~al.\ 2003) would
place the (initial?) reionization at $z \sim 10$--20.

A simple picture of a clean-cut, universally synchronized
Gunn-Peterson trough now appears unlikely. The key issue is the
clumpiness of the IGM, and the gradual development and clumpy
distribution of the first ionizing sources, either protogalaxies or
early AGN (see, e.g., Miralda-Escude et~al.\  2000).  
Reionization is expected to occur gradually as the UV emissivity
increases (cf.\ \mbox{McDonald} \& Miralda-Escude 2001) with the lowest
density regions becoming fully reionized first.  This is also
suggested by modern numerical simulations (e.g., Gnedin \& Ostriker
1997; Ciardi et~al.\ 2003; Gnedin 2004) which predict an
extended period of reionization, starting at $z \sim 15$ or even
higher and ending at $z \sim 6$ or so (see also Wyithe \& Loeb 2003, Cen 2003,
Haiman \& Holder 2003, Somerville et~al.\ 2003).

Large samples of normal galaxies are now studied out to $z \sim 4.5$
(Steidel et~al.\ 1999, Dawson et~al.\ 2004).  
Surveys like, e.g., GOODS (Giavalisco
et~al.\ 2004, Dickinson et~al.\ 2004) and HUDF (Beckwith
et~al.\ in preparation; see also Bunker et~al.\ 2004, Yan \&
Windhorst 2004), have provided us with the first significant samples
of galaxies at $z \gta 5$.  However, if our understanding of galaxy
formation through hierarchical clustering is correct, we expect the
earliest objects to be rare and highly clustered rather than typical
field galaxies. Indeed, some evidence of clustering at redshift around
6 has been found in the UDF (Malhotra et~al.\ 2004, see also Ouchi 
et~al.\ 2005).  This leads us naturally to the study of $z\gta6$ QSO
environments as they should be the most clustered environments at
these very high redshifts.

Indeed, a generic expectation in most models of galaxy formation is
that the most massive density peaks in the early universe are likely
to be strongly clustered (Kaiser 1984; Efstathiou \& Rees 1988).  The
evidence for such bias is already seen in large samples of
Lyman-break galaxies at $z \sim 3$--3.5 (Steidel et~al.\  2003),
and in Lyman $\alpha$ selected galaxy samples (e.g., Venemans 
et~al.\ 2003); it should be even stronger at higher redshifts.  An
``excess'' in the number of galaxies and in the density of star
formation was also discovered in a systematic Keck survey of fields
centered on known $z > 4$ quasars (e.g., Djorgovski 1999, Djorgovski
et~al.\ 1999, 2003, and in prep.).  The high metallicity
associated with QSOs---even at $z \gta 6$---is often interpreted as
evidence that they are located at the center of massive
(proto--)galaxies, thereby corroborating the overall picture.  These
arguments justify the expectation that QSOs at $z\simeq6$ most likely
highlight some of the first perturbations that become non--linear in
the density distribution of matter.

This paper is devoted to the study of the environment of the QSO SDSS
J1030+0524. Section~2 describes the data and our data reduction
techniques. Section~3 describes the catalogs and the candidate list.
Section~4 presents a spectrum obtained for one
of our candidates. Section~5 presents and discusses our results.

\section{Data reduction and analysis}

We have observed a field near SDSS J1030+0524 using the Advanced
Camera for Surveys (ACS) onboard the Hubble Space Telescope. The Wide
Field Camera (WFC) was used to secure an integration of 5,840~s in
F775W (hereafter $i_{775}$) and of 11,330~s in F850LP (hereafter
$z_{850}$). The relative exposure times in these filters have been
chosen so as to have a depth in these two filters comparable to that
of the GOODS survey (Giavalisco et~al.\ 2004) allowing us to use GOODS
as a reference field sample. Since SDSS J1030+0524 is about 30 arcsec from
a 12th magnitude star, we placed the star in the interchip gap of the WFC. This 
avoids any bleeding along the CCD columns.
The images were dithered to improve the PSF sampling and we
obtained a total of 7~images in F775W and 9~images in F850LP.

The data were processed by the ACS pipeline (CALACS), which performs the
basic data reduction steps of bias subtraction, dark subtraction and flat
fielding of the raw data, resulting in the flat-fielded images (the {\it flt\/} files).
The post-pipeline processing steps include the creation of individual
weight maps for each {\it flt\/} file, drizzling to the same scale as
GOODS (0.03 arcsec/pixel) using MultiDrizzle (Koekemoer et~al.\ 2002),
correcting for correlated noise introduced by MultiDrizzle, and masking
any bad pixels and pixels that were less than 10\% of the total
exposure time. The final MultiDrizzle run was performed with parameters 
$\mathit{pixscale}=0.6$ and $\mathit{pixfrac}=1.0$ and with 
$\mathit{final}\_\mathit{wht}\_\mathit{type}=\mathit{ivm}$ to make use of 
individual weight maps. Special care was taken in producing the weight maps 
used as an input to MultiDrizzle, which also helps eliminate many spurious 
detections from the resulting object catalogs.

Source detection was done on the $z_{850}$-band images, using the same 
version of Sextractor (Bertin \& Arnout 1996) used by the GOODS collaboration 
and the same input parameters in order to facilitate a straightforward
comparison between the two catalogs.  The software was run with the
MultiDrizzle output image ({\it drz\/} file) and weight map (converted
to an rms map). The Sextractor parameters used attempt to maximize the
number of faint sources found while minimizing the number of false
detections. We performed isophotal photometry, mag\_auto photometry,
photometry with a 16 pixel fixed diameter and photometry at the half
light radius. The photometric errors calculated by Sextractor are
measured from the weight maps and take into account the background and
instrumental noise and exposure time information.  We verified that
the various photometric measurement were consistent and for the rest
of the analysis we used the mag\_auto photometry. We have only
considered sources detected in $z_{850}$ at $S/N\geq5$. For these
sources we have computed the $(i_{775}-z_{850})$ color directly from
the $i_{775}$ magnitude when the $i_{775}$ flux was measured to better
than 2$\sigma$ or otherwise from the 2$\sigma$ upper limit to the
$i_{775}$ flux.  We have run completeness simulations in the SDSS 
J1030+0524 field and found that our photometry has comparable 
completeness to GOODS. We find that we reach 50\% completeness 
for compact galaxies at $z_{850}\simeq 27$. Finally, we have verified 
that our catalog contains no contamination by noise peaks mistakenly 
identified as galaxies by running a negative image test (Dickinson 
et~al.\ 2004, Yan \& Windhorst 2004).

\section{Candidate objects at $\lowercase{z}=6$}

The final Sextractor catalog contains 1551 objects. Fig.~1a shows the
counts for the SDSS J1030+0524 field and those for the GOODS N and S fields
corrected to the same area. The QSO is recovered as the reddest object
in the field. Fig.~1b shows the color distribution of all objects with
$z\leq26.5$. The basic distribution in both counts and color are similar
in the SDSS J1030+0524 area and in GOODS. This is an independent test that
the GOODS catalog and ours are generally consistent.

Excluding the QSO, a total of 8 objects have $(i_{775}-z_{850})>1.3$,
$S/N > 5$ and $z_{850} \leq 26.5$. We expect that a high fraction of
galaxies with $(i_{775}-z_{850})>1.3$ will be at $z \gta 5.5$ as
Malhotra et~al.\ (2005) have shown that at 
$z_{850}\leq27$ about 90\% of these objects are truly at high-redshift.
At our color cut, and in agreement with Willott et~al.\ (2004),
we do not find a significant overall source density excess between the
SDSS J1030+0524 field and GOODS. However, when we adopt the more
conservative cut $(i_{775}-z_{850})>1.5$, we still find 7 objects (excluding
the QSO) in the SDSS J1030+0524 field against the 3.33 expected from
GOODS. This is illustrated graphically in Fig.~1c where we show the
counts of objects redder than $(i_{775}-z_{850})=1.5$. 
The objects are generally resolved as shown in Fig.~2 for two
objects from this selection.

\section{A First Attempt at Spectroscopic Confirmation}

Spectroscopic observations of eight of the color-selected candidates were
obtained using the LRIS instrument (Oke et~al.\ 1995) on the W.M.\ Keck
Observatory's 10-m Keck-I telescope in multislit mode.  The data were
obtained on the nights of UT 2004 December 10 and 15, in moderate to
good conditions, with an effective total exposure time of 2.8 hours
per night.  All observations employed a 400 lines mm$^{-1}$ grating
($\lambda_{\rm blaze} = 8500$~\AA), giving an effective resolution of
FWHM $\sim 8$~\AA.  The data were reduced using standard procedures and
flux-calibrated using archival data obtained in photometric conditions
through the same instrument configuration.
                                                                           
Only one of the targeted sources, J103024.08+ 052420.4, shows an emission
line, allowing an unambiguous redshift identification.  Its spectrum,
derived from the UT 2004 December 10 data when conditions were more
favorable, is shown in Fig.~3.  The emission line reproduced on UT 2004
December 15.  The spectrum shows a single, isolated, asymmetric, high
equivalent width emission line, characteristic of high-redshift Ly$\alpha$
emission (cf.\ Stern et~al.\ 2005).  A weak continuum decrement across
the emission line is also evident, characteristic of absorption due to
the Ly$\alpha$ forest at high redshift.  We derive a redshift of $z =
5.970$ for this source, confirming that our $i$-drop criteria reliably
identify $z \sim 6$ galaxies.
                                                                           
The spectra of the remaining targets do not have a sufficient
signal-to-noise to make any other firm identifications at this point.
Further analysis of these and other, related data will be presented
elsewhere.

\section{Discussion and Conclusions}

The probability of finding 7 objects when one expects 3.33 is 0.7\% if
the distribution is Poissonian. However, the true probability would
probably be higher because of clustering. Indeed, by randomly
overlaying WFC fields within GOODS we find that the probability
of one such field containing 7 objects with $(i_{775}-z_{850})\geq1.5$
is about 3\%.

Luckily we do not only have the number of red objects but also their
distribution in color and this also turns out to be different from
that within GOODS. In Fig.~1d we show the cumulative distribution of
objects within the SDSS J1030+0524 field and in GOODS (renormalized to the
same area). Clearly, the SDSS J1030+0524 object distribution appears to be
different from that of GOODS. We have applied a Kolmogorov-Smirnov
test to determine whether the distribution of the 8 sources in
SDSS J1030+0524 redder than $(i_{775}-z_{850})=1.3$ and that of the 252 in
GOODS with the same color cut are drawn from the same parent
distribution. The result is a difference value of $D=1.72$
corresponding to a probability 0.55\%.

Thus, we find that the SDSS J1030+0524 field appears to have an excess of
red objects. Our result is in agreement with Willott et~al.\ (2004) as our data 
set is deeper and the majority of the excess sources are fainter than 
$z_{850}=25.5$.

If the SDSS J1030+0524 does indeed have an excess of red sources, one
may wonder whether their redshift is the same as the QSO. The selection 
$(i_{775}-z_{850})\geq1.5$ should be effective at selecting objects at 
redshift greater than 5.9. Indeed, the one object for which we could obtain 
a spectroscopic confirmation satisfies this constraint. Most of the sources 
are too faint to be able to measure a color as red as that of the QSO but 3 
of the 7 sources are detected as dropouts, i.e. their $i_{775}$ band flux is 
measured at less than 2$\sigma$.  It should also be noted that SDSS QSOs 
at $z>6$ are extremely rare objects and are presumably associated to rare 
density peaks which may correspond to perturbations on very large scales. 
The size of large scale structures seen at lower redshift can go up to at least 
50 comoving Mpc which would correspond to $\delta z = 0.12$ at $z=6.28$. 
Thus, we expect some, but not all, of the excess sources to be at $z=6.28$. 
We do not know whether J103024.08+052420.4 belongs to the same 
overdensity as the QSO as the redshift difference is sizeable. However, one 
can speculate that if it did, it would imply a minimum size for the overdensity
of $\sim 130$ comoving Mpc along the line of sight. If the overdensity had 
this typical size also in the transverse direction, this would translate into an 
angular extension of $\sim 50$ arcmin.

The discovery of a density excess near SDSS J1030+0524 is potentially
very interesting and requires a further confirmation.  Additional spectroscopic
follow-up is now in progress. The presence of a very extended overdensity could
be verified by repeating this search over a much larger area. We also plan
to repeat this study on 4 more fields around redshift $\sim6$ QSOs,
which will be reported elsewhere.

\acknowledgments

We thank the GOODS teams for obtaining and releasing the data on which
our study is partly based. NASA/ESA HST data are obtained at STScI,
which is operated by AURA, Inc., under NASA contract NAS5-26555.  This
work is partially supported by NASA Grant GO-9777. We thank the staff of
the W.M.\ Keck Observatory for their expert assistance. The work of DS was
carried out at the Jet Propulsion Laboratory, California Institute of Technology, 
under a contract with NASA.

\begin{figure}[h]
\epsscale{.60}
\plotone{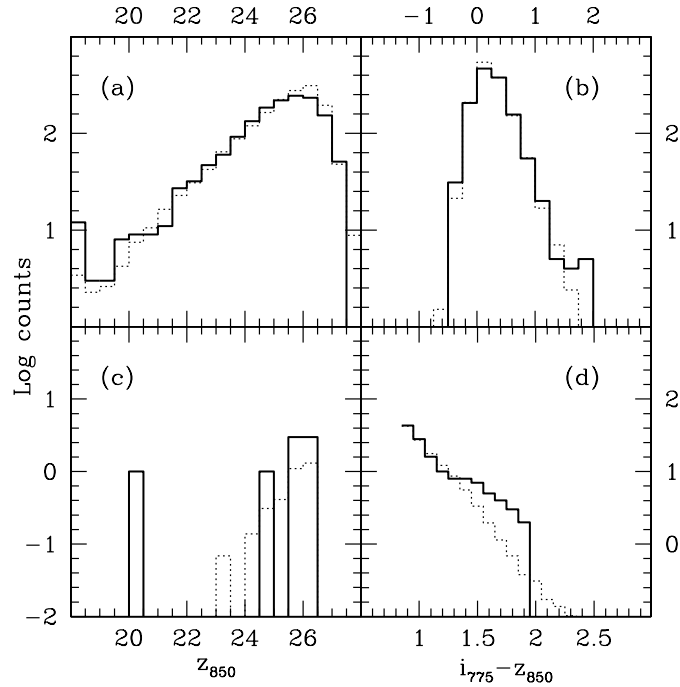}
\caption{Comparison of sources in the field around SDSS J1030+0524 (thick
lines) with those in GOODS (thin lines). The GOODS counts have been
renormalized by the ratios of the area of the two surveys and only
sources with $S/N \geq 5$ are shown.  Panel~(a) shows the total
counts. All remaining panels include only objects brighter than
$z_{850}= 26.5$.  Panel~(b) shows the color distribution within the
two catalogs. Both panels~(a) and~(b) illustrate how the generic
properties of objects in the two catalogs are comparable. Panel~(c)
shows the counts of objects redder then $i_{775}-z_{850} = 1.5$. 
Panel~(d) shows the number of objects redder than a given
$i_{775}-z_{850}$ as a function of $i_{775}-z_{850}$. In panel~(d) we
have omitted 4 objects that didn't look convincing by visual
inspection (no similar check was done on the GOODS sources). The field
of SDSS J1030+0524 appears to contain an excess of sources with
$i_{775}-z_{850}\geq 1.5$.}
\end{figure}

\begin{figure}
\epsscale{.8}
\plotone{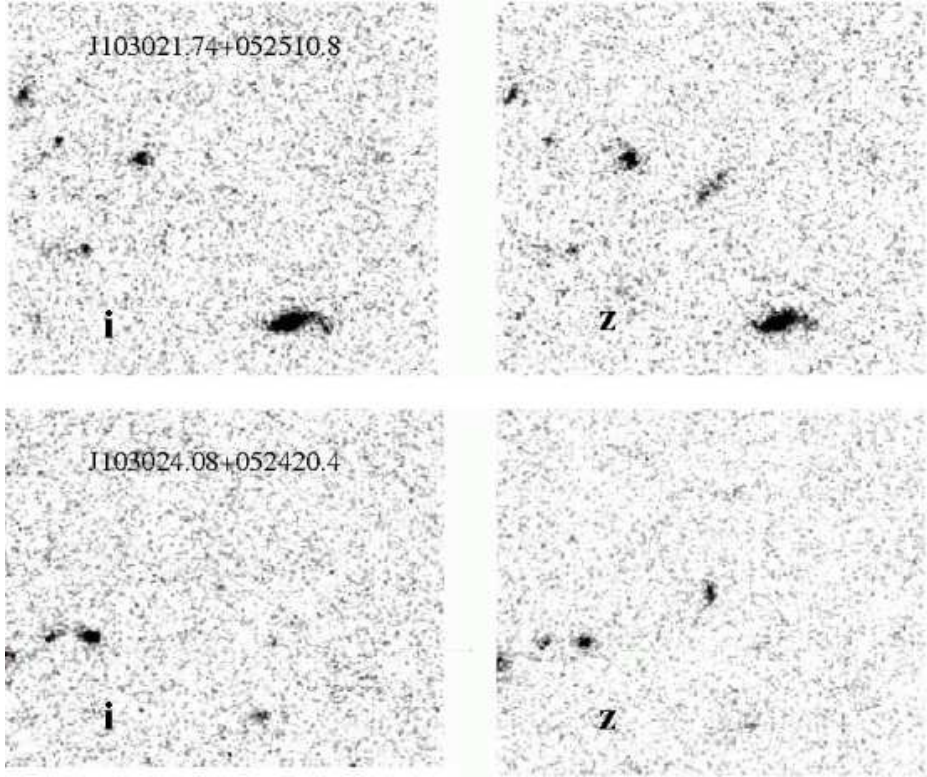}
\caption{Images of two of the reddest objects in the SDSS J1030+0524 field:
J103021.74+052510.8 (top) and J103024.08+052420.4 (bottom). For each
we show the image in F775W on the left and the one in F850LP on
the right. Both objects appear resolved in the F850LP images.}
\end{figure}

\epsscale{.8}
\begin{figure}
\plotone{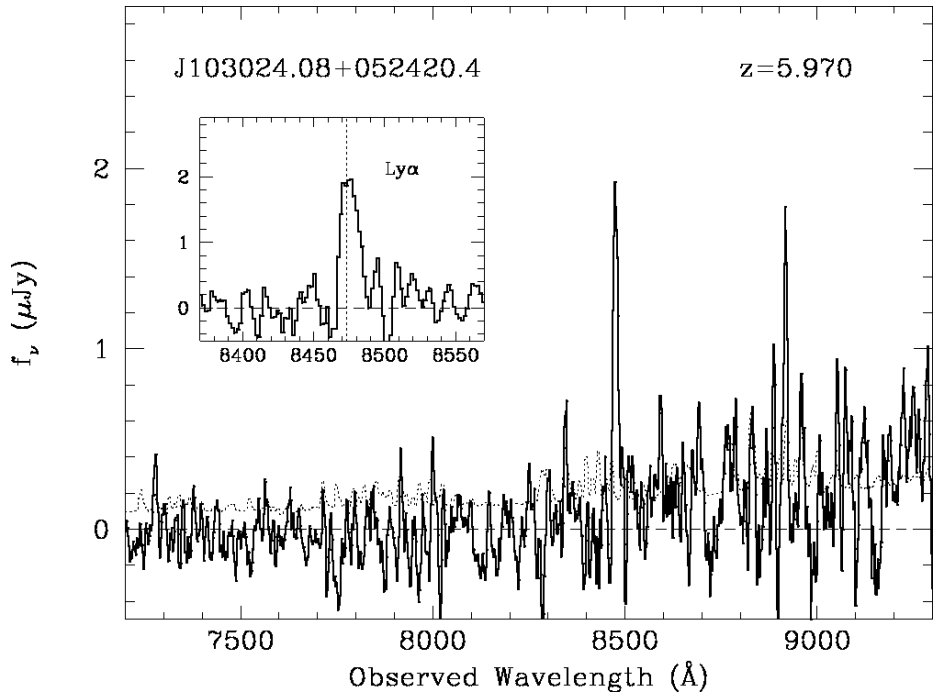}
\caption{Spectrum of J103024.08+052420.4 at $z = 5.970$, obtained with LRIS on the
Keck-I telescope.  The total exposure time is 2.8~hr, and the spectrum was
extracted using a 1\farcs2 $\times$ 1\farcs5 aperture.  The spectrum has
been smoothed using a 5.5~\AA\, boxcar filter.  The dotted spectrum in the
main panel indicates the corresponding 1$\sigma$ error spectrum, assuming
Poisson fluctuations of the sky spectrum. }
\end{figure}


\begin{thebibliography}{}
\bibitem[Barkana and Loeb (2001)]{barkana01} Barkana, R., \& Loeb, A. 2001,
Phys.\ Rep., 349, 125
\bibitem[Becker et al.\ (2001)]{becker01} Becker, R.~H., et~al. 2001, \aj, 122,
2850
\bibitem[Bertin \& Arnouts (1996)]{bertin} Bertin, E. \& Arnouts, S. 1996, A\&AS, 117, 393
\bibitem[Bunker et al.\ (2004)]{bunker04} Bunker, A.~J., Stanway, E.~R., Ellis, R.~S. \& McMahon, R.~G. 2004, \mnras, 355, 374
\bibitem[Cen (2003)]{cen03} Cen, R. 2003, \apj, 591, 12
\bibitem[Ciardi et al.\ (2003)]{ciardietal03} Ciardi, B., Ferrara, A., \& White, S.~D.~M. 2003, \mnras, 344, L7
\bibitem[Dawson et al.\ (2004)]{dawson04} Dawson, S., et~al. 2004, ApJ, 617, 707
\bibitem[Dickinson et al.\ (2004)]{dickinsonz6} Dickinson, M., et~al. 2004, \apjl, 600, L99
\bibitem[Djorgovski et al.\ (1999)]{djorg+99} Djorgovski, S.~G., Odewahn, S.~C., Gal, R.~R., Brunner, R., \& de Carvalho, R. 1999, ASPCS 191, 179
\bibitem[Djorgovski (1999)]{djorgo99} {} Djorgovski, S.~G. 1999, ASPCS, 193, 397
\bibitem[Djorgovski et al.\ (2001)]{djorgo01} Djorgovski, S.~G., Castro, S., Stern, D., \& Mahabal, A. 2001, \apjl, 560, L5
\bibitem[Djorgovski et al.\ (2003)]{djorgo03} Djorgovski, S.~G., Stern, D., Mahabal, A.~A. \& Brunner, R. 2003, \apj, 596, 67
\bibitem[Djorgovski (2003)]{djorgo04} Djorgovski, S.~G. 2004, preprint (astro-ph/0409378)
\bibitem[Efstathious \& Rees (1988)]{efstathiourees88} Efstathiou, G. \& Rees, M.~J. 1988, \mnras, 235, 715
\bibitem[Fan et al.\ (2001)]{fan01} Fan, X., et~al. 2001, \aj, 122, 2833
\bibitem[Giavalisco et al.\ (2004)]{giavaliscointro} Giavalisco, M., et~al. 2004, \apjl, 600, L93
\bibitem[Gnedin (2000)]{gnedin00} Gnedin, N.~Y. 2000, \apj, 542, 535
\bibitem[Gnedin (2004)]{gnedin04} Gnedin, N.~Y. 2004, \apj, 610, 9
\bibitem[Gnedin and Ostriker (1997)]{gnedinostriker97} Gnedin, N.~Y., \& Ostriker, J.~P. 1997, \apj, 486, 581
\bibitem[Gunn and Peterson (1965)]{gunn65} Gunn, J.~E., \& Peterson, B.~A. 
1965, \apj, 142, 1633
\bibitem[Haiman \& Holder (2003)]{haimanholder03} Haiman, Z. \& Holder, G.~P. 
2003, \apj, 595, 1
\bibitem[Loeb \& Barkana (2002)]{loeb02} Loeb, A. \& Barkana, R. 2002, ARAA, 39, 19
\bibitem[Kaiser (1984)]{kaiser84} Kaiser, N. 1984, \apjl, 284, L9
\bibitem[Koekemoer et al.\ (2002)]{koeke02} Koekemoer, A.~M., Fruchter, A.~S., Hook, R. \& Hack, W. 2002, HST Calibration Workshop, STScI, Arribas, et~al.\ eds, 337
\bibitem[Kogut et al.\ (2003)]{kogut03} Kogut, A. et~al.\ (the WMAP team)\ \ 2003, \apjs, 148, 161
\bibitem[Madau (2000)]{madau00} Madau, P., Philosophical Transactions of the Royal Society of London, Series~A, Vol.~358, 
no.~1772, 2021
\bibitem[McDonald \& Miralda-Escud{\'e} (2001)]{mcdonald01} McDonald, P., \& Miralda-Escud{\'e}, J. 2001, \apjl, 549, L11 
\bibitem[Miralda-Escud{\'e} et al.\ (2000)]{miraldaetal00} Miralda-Escud{\'e}, J., Haehnelt, M., \& Rees, M~J. 2000, \apj, 530, 1
\bibitem[Oke et al.\ (2005)]{oke2005} Oke, J.~B., et~al. 1995, PASP, 107, 375
\bibitem[Ouchi et al.\ (2005)]{ouchi2005} Ouchi, M., et~al. 2005, \apjl, 620, L1
\bibitem[Somerville et al.\ (2003)]{somerville03} Somerville, R.~S., Bullock, J.~S. \& Livio, M. 2003, \apj, 593, 616
\bibitem[Steidel et al.\ (1999)]{steidel99} Steidel, C.C., Adelberger, K.~L, Giavalisco, M., Dickinson, M., \& Pettini, M.
1999, \apj, 519, 1
\bibitem[Stern et al.\ (2005)]{stern05} Stern, D., et~al. 2005, ApJ, 619, 12
\bibitem[Steidel et al.\ (2003)]{steidel03} Steidel, C.~C., Adelberger, K.~L., Shapley, A.~E., Pettini, M., Dickinson, M. \& Giavalisco, M. 2003, \apj, 592, 728
\bibitem[Stiavelli (2002)]{stia02} Stiavelli, M. 2002, in ``Future Research Directions and Visions for Astronomy," SPIE, A. Dressler Ed., p.~122
\bibitem[Venemans et al.\ (2003)]{venemans03} Venemans, B.~P., Kurk, J.~D., Miley, G.~K. \& R{\"o}ttgering, H.~J.~A. 2003, New 
Astr.\ Rev., 47, 353
\bibitem[White et al.\ (2003)]{white03} White, R., Becker, R., Fan, X., \& Strauss, M. 2003, \aj, 126, 1
\bibitem[Willott et al.\ (2004)]{willott04} Willott, C.~J.,  et~al. 2004, in ``Growing Black Holes,'' ESO (Garching), Merloni, et~al.\ eds, in press (astro/ph 0410306)
\bibitem[Wyithe and Loeb (2003)]{wyitheloeb03b} Wyithe, J.~S.~B., \& Loeb, A. 2003, \apjl, 588, L69
\bibitem[Yan \& Windhorst (2004)]{yanwind04} Yan, H. \& Windhorst, R.~A. 2004, \apjl, 612, L93
\end{thebibliography}
\end{document}